\documentstyle[12pt,epsf]{article}

\newcommand{\AH}{{\cal N}_{\rm AH}}
\newcommand{\TN}{{\cal N}_{\rm TN}}
\newcommand{\EH}{{\cal N}_{\rm EH}}

\newcommand{\Z}{{\bf Z}}

\newcommand{\lab}[1]{\label{#1}}

\newcommand{\ol}{\overline}
\def\Tr{\mathop{\rm Tr}\nolimits}
\newcommand{\el}[1]{\makebox[0.7em]{$#1$}}

\newcommand{\J}[4]{{\sl #1} {\bf #2} (#3) #4}
\newcommand{\NP}{Nucl.\ Phys.}
\newcommand{\PL}{Phys.\ Lett.}

\begin{document}

\baselineskip=18pt plus 0.2pt minus 0.1pt

\begin{titlepage}
\title{
\hfill\parbox{4cm}
{\normalsize KUNS-1468\\HE(TH)~97/15\\{\tt hep-th/9710026}}\\
\vspace{1cm}
D-particle creation on an orientifold plane
}
\author{
Yosuke Imamura\thanks{{\tt imamura@gauge.scphys.kyoto-u.ac.jp}}
{}\thanks{
Supported in part by Grant-in-Aid for Scientific
Research from Ministry of Education, Science and Culture
(\#5416).}
\\[7pt]
{\it Department of Physics, Kyoto University, Kyoto 606-01, Japan}
}
\date{}

\maketitle
\thispagestyle{empty}

\begin{abstract}
\normalsize
We study the propagations of gravitational wave
and D-particle on D6-brane
and orientifold 6-plane backgrounds in the M-theory framework.
In the case of orientifold plane, D-particle number is not conserved
and gravitational wave can convert into D-particle.
For the simplest case, we calculate its amplitude numerically.
\end{abstract}

\end{titlepage}

\section{Introduction}
Recently, D6-brane and orientifold 6-plane are
of great interest\cite{Sen42,SenEnhance,SigmaModel,SenAug,Imamura}
 in M-theory\cite{WittenM}.
These $6+1$ dimensional objects in type IIA theory are expressed
as eleven dimensional smooth manifolds in M-theory,
while other brane solutions generally have a singularity.
Therefore, in low energy and strong coupling limit,
their dynamics can be analyzed by means of eleven dimensional supergravity
without informations of its microscopic physics.

The manifolds which express the D6-brane (D6) and
the orientifold 6-plane (O6) are
${\cal N}_{\rm TN}\times M_7$ and ${\cal N}_{\rm AH}\times M_7$,
respectively\cite{SenEnhance,SW96},
where ${\cal N}_{\rm TN}$ is Taub-NUT manifold,
${\cal N}_{\rm AH}$ is Atiyah-Hitchin manifold\cite{AH}
and $M_7$ is the $6+1$-dimensional flat Minkowski space.
The D6-brane lying on the orientifold 6-plane (O6+D6) corresponds to
$\ol{\cal N}_{\rm AH}\times M_7$, where $\ol{\cal N}_{\rm AH}$ is
covering space of ${\cal N}_{\rm AH}$\cite{SenAug}.

The purpose of this paper is to study the propagations
of the gravitational wave or the D-particle
on D6, O6 or O6+D6 backgrounds in type IIA theory with strong coupling.
We calculate the potential
between a 6-plane and a D-particle and reproduce the known result.
In the case of the O6 or the O6+D6 background,
we find the process
in which gravitational wave colliding with the fixed plane
convert into a D-particle.
For eigenstates of orbital angular momentum,
we calculate the transition amplitude of this process
numerically in the simplest case.
As a result, we obtain the amplitude of order one.

\section{Review of hyper K\"ahler manifold with $SU(2)$ isometry}
The manifolds $\EH$, ${\cal N}_{\rm TN}$, ${\cal N}_{\rm AH}$
and $\ol{\cal N}_{\rm AH}$
have some common properties
(${\cal N}_{\rm EH}$ is Eguchi-Hanson manifold,
which is not used in this paper).
Namely, they are hyper K\"ahler manifolds with $SU(2)$
(or $SO(3)$) isometry.
Such manifolds are studied in detail in \cite{AH}
and metrics of these manifolds are given explicitly.
In this section, we review the properties of these manifolds
which are necessary in the later sections.

Topologically, these manifold can be expressed
as ${\cal N}={\cal M}/\Gamma\times R^+$,
where $\cal M$ is a group manifold of $SU(2)$
and $R^+$ is a half line (Fig.\ref{xislm}).
$\Gamma$ is a discrete subgroup of $SU(2)$ which
we shall mention at the end of this section.
At the end point of $R^+$, $\cal M$ shrinks
to a point (${\cal N}_{\rm TN}$)
or two sphere (${\cal N}_{\rm EH}$, ${\cal N}_{\rm AH}$
and $\ol{\cal N}_{\rm AH}$).
\begin{figure}[hhh]
\epsfysize=4cm
\begin{center}
\leavevmode
\put(85,5){\large$R^+$}
\put(95,105){\large$\cal M$}
\put(170,10){\large$\rho$}
\put(-40,55){\large$\rho=0$}
\epsfbox{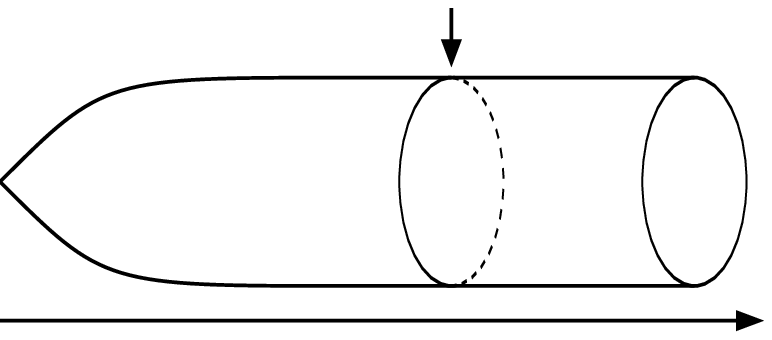}
\caption{manifold $\cal N$}
\lab{xislm}
\end{center}
\end{figure}
Each point on $\cal N$ is specified by a pair $(\rho,g)$
of real number $\rho\geq0$ and an element $g\in SU(2)$,
which parameterize $R^+$ and $\cal N$, respectively.

To determine the $SU(2)$ invariant metric on $\cal N$,
we introduce two mappings $SU(2)_L$ and $SU(2)_R$
from $\cal N$ onto itself,
\begin{eqnarray}
SU(2)_L&:&(\rho,g)\rightarrow(\rho,hg),\quad h\in SU(2),\\
SU(2)_R&:&(\rho,g)\rightarrow(\rho,gh),\quad h\in SU(2),
\end{eqnarray}
and parameterize the tangent space
$T\cal M$ at $g\in\cal M$
by the local coordinate $\epsilon^{1,2,3}$ defined by
\begin{equation}
U^{(s)}(g+dg)=U^{(s)}(g)\exp\left(T^{(s)}_a\epsilon_a\right),
\end{equation}
where $U^{(s)}(g)$ is a representation matrix and 
$T_a^{(s)}$ is the generator of $SU(2)$ normalized
as $[T^{(s)}_a,T^{(s)}_b]=-\epsilon_{abc}T^{(s)}_c$.
The index $s$ specifies the representation of $SU(2)$.
Because $SU(2)_R$ and $SU(2)_L$ are commutable each other,
this local coordinate system defined by $SU(2)_R$
is invariant under the $SU(2)_L$ rotation.
Therefore, if we take the $g$-independent metric
$ds^2=g_{ab}(\rho)\epsilon^a\epsilon^b$
in this local coordinate,
the rotation $SU(2)_L$ is an isometry of $\cal M$.
Adding the radial parameter $\rho$, which we define as the geodesic distance
from the center $\rho=0$,
the metric on $\cal N$ is given by
\begin{equation}
ds^2=d\rho^2+g_{ab}(\rho)\epsilon^a\epsilon^b.
\end{equation}
We can diagonalize the metric $g_{ab}$ by redefining $\epsilon^a$:
\begin{equation}
g_{ab}(\rho)=\left(\begin{array}{ccc}
a^2(\rho)\\
&b^2(\rho)\\
&&c^2(\rho)
\end{array}\right).
\lab{diagonal}
\end{equation}
The three functions $a(\rho)$, $b(\rho)$ and $c(\rho)$ give
the moving distances of the mappings $U(1)_x,U(1)_y,U(1)_z\subset SU(2)_R$,
which are generated by $T_x$, $T_y$ and $T_z$, respectively.


For the manifold $\cal N$ to be a hyper K\"ahler manifold,
the connection should be anti-self-dual in some coordinate.
This requires the functions $a(\rho)$, $b(\rho)$ and $c(\rho)$
to satisfy the equations
\begin{eqnarray}
\frac{da}{d\rho}=\frac{-a^2+b^2+c^2}{2bc}-k,\quad
\frac{db}{d\rho}=\frac{-b^2+c^2+a^2}{2ca}-k,\quad
\frac{dc}{d\rho}=\frac{-c^2+a^2+b^2}{2ab}-k,
\lab{CondABC}
\end{eqnarray}
where $k=0$ for ${\cal N}_{\rm EH}$ and $k=1$
for ${\cal N}_{\rm TN}$, ${\cal N}_{\rm AH}$ and $\ol{\cal N}_{\rm AH}$.

The functions $a(\rho)$, $b(\rho)$ and $c(\rho)$ for the manifolds
${\cal N}_{\rm TN}$, ${\cal N}_{\rm AH}$ and $\ol{\cal N}_{\rm AH}$,
in which we are interested in this paper,
have the common behavior in the asymptotic region ($\rho\rightarrow\infty$):
\begin{equation}
|a(\rho)|\sim|b(\rho)|\sim\rho,\quad
|c(\rho)|\sim c_0\pm\frac{c_0^2}{2\rho},\quad
\mbox{( $+$ : ${\cal N}_{\rm AH}$, $\ol{\cal N}_{\rm AH}$,
$-$ : ${\cal N}_{\rm TN}$)}.
\lab{asym}
\end{equation}
This behavior implies that the structure of $\cal N$ in the asymptotic region
is $R^3\times S^1$, with $S^1$ being the orbit of $U(1)_z$
which will be identified
with the $X^{11}$ direction of M-theory.
The radius of this $S^1$ is given by $2c(\rho)$.
Because $U(1)_z$ mixes the two directions $\epsilon^1$ and $\epsilon^2$,
it becomes an isometry of $\cal N$
when $a^2(\rho)=b^2(\rho)$.
In the asymptotic region, all of $\TN$, $\AH$
and $\ol\AH$ have this isometry,
while it is broken in the central region for $\AH$
and $\ol\AH$.

If eq.(\ref{CondABC}) is satisfied, $\cal N$
is a hyper K\"ahler manifold with $SU(2)$
isometry, except for the point or the two-sphere at $\rho=0$.
Generically, this point or two-sphere is a conical singularity.
To avoid such a singularity, we should divide the manifold $\cal M$
by a discrete subgroup $\Gamma\subset SU(2)_R$,
which is given in the table \ref{table}
($D=\{\pm1,\pm i\sigma_x,\pm i\sigma_y,\pm i\sigma_z\}$).
\begin{table}[hhh]
\begin{center}
\begin{tabular}{|c|cccc|}
\hline
manifold & ${\cal N}_{\rm TN}$ & ${\cal N}_{\rm EH}$ &
 ${\cal N}_{\rm AH}$ & $\ol{\cal N}_{\rm AH}$ \\
\hline
$\Gamma$ & $\{e\}$ & $Z_2$ & $D$ & $Z_4$ \\
\hline
$\Gamma\cap U(1)_x$ & $\{e\}$ & $\{e\}$ & $Z_4$ & $Z_4$ \\
\hline
$\Gamma\cap U(1)_y$ & $\{e\}$ & $\{e\}$ & $Z_4$ & $Z_2$ \\
\hline
$\Gamma\cap U(1)_z$ & $\{e\}$ & $Z_2$ & $Z_4$ & $Z_2$ \\
\hline
\end{tabular}
\end{center}
\caption{}
\lab{table}
\end{table}
We mentioned before that
the compactification radius is given by $2c(\rho)$.
However, we should take account of the action of $\Gamma$ on this $S^1$:
this circle should be divided by the discrete group $\Gamma\cap U(1)_z$.
As a result, we obtain the correct relation between
$c(\rho)$ and compactificaton radius $R_{11}$,
\begin{equation}
R_{11}=\frac{2}{p}c(\rho),\lab{R11is}
\end{equation}
where $p=1,2,4$ for ${\cal N}_{\rm TN}$,
$\ol{\cal N}_{\rm AH}$ and ${\cal N}_{\rm AH}$, respectively.
\section{Mode Expansion of Scalar Field on $\cal N$}
The bosonic fields in eleven dimensional supergravity
are the metric $G_{MN}$ and the three form field $A_{LMN}$.
To study the D-particle or gravitational wave scattering
by D6, O6 or O6+D6 in M-theory,
we should analyze the propagation of these fields
on the manifold ${\cal N}\times M_7$
where $\cal N$ is one of ${\cal N}_{\rm TN}$, ${\cal N}_{\rm AH}$
and $\ol{\cal N}_{\rm AH}$.
We label the directions of $M_7$ by $0,1,\ldots,6$
and that of $\cal N$ by $7,8,9,11$.
Since the wave propagation along the $M_7$ direction is trivial,
the problem is reduced to solving the wave equation on the manifold $\cal N$.
On this four dimensional manifold,
$G_{MN}$ and $A_{LMN}$
with $L,M,N=0,\ldots,6$ are regarded as scalar fields.
Once the problem is solved for these scalar fields,
we can get the solution for other modes (vector, tensor and spinor)
by means of the supersymmetry.
Therefore we may focus only on the scalar modes.
The problem we consider in this section is mathematically identical to
what is investigated in detail by N.S.Manton et al.
in the context of monopole scattering\cite{Manton,GibbonsManton}.

It is known that we can construct
the complete system of complex scalar functions
on the group manifold
from the representation matrices $U^{(s)}(g)$ as follows:
\begin{equation}
\phi_{s,m,n}(g)=U^{(s)}_{m,n}(g),
\end{equation}
where the index $s$ labels the representations of the group.
In the ${\cal M}=SU(2)$ case we shall consider,
$s$ is an integer or a half integer, and $m$ and $n$ satisfy
\begin{equation}
-s\leq m,n\leq s,\quad
s-m,s-n\in\Z.
\end{equation}
It is worth giving the physical interpretation of these indices here.
The indices $s$ and $m$ are quantum numbers for $SU(2)_L$.
Because $SU(2)_L$ is identified
with the rotation of $X^{7,8,9}$ in the target space,
these indices represent the orbital angular momentum
of the D-particle or the gravitational wave.
The index $n$ is the charge of $U(1)_z\in SU(2)_R$
(i.e., the shift along $S^1$ of the $X^{11}$ direction),
and therefore it is regarded as the D-particle number.
From eq.(\ref{R11is}), the relation between $n$
and the D-particle number $N_D\in\Z$ is
\begin{equation}
N_D=\frac{2}{p}n.\lab{NDis}
\end{equation}

Any scalar function $\phi(\rho,g)$ on $\cal N$ is expanded as
\begin{equation}
\phi(\rho,g)=\sum_{s,m,n}c^{(s)}_{nm}(\rho)U^{(s)}_{mn}(g)
            =\sum_s\Tr(c^{(s)}(\rho)U^{(s)}(g)),
\lab{expand}
\end{equation}
and its derivative with respect to local coordinate $\epsilon^a$ is
\begin{eqnarray}
\partial_a\sum\Tr(c^{(s)}(\rho)U^{(s)}(g))
&=&\partial_a\sum_s\Tr[c^{(s)}(\rho)U^{(s)}(g)
       \exp(T^{(s)}_a\epsilon^a)]|_{\epsilon^a=0}\nonumber\\
&=&\sum_s\Tr[c^{(s)}(\rho)U^{(s)}(g)T^{(s)}_a].
\end{eqnarray}
Laplacian on the manifold $\cal M$ is given by
\begin{eqnarray}
\Delta^{\cal M}\phi(\rho,g)
&=&g^{ab}\partial_a\partial_b\sum_s\Tr(c^{(s)}(\rho)U^{(s)}(g))\nonumber\\
&=&g^{ab}\sum_s\Tr\left[c^{(s)}(\rho)\{U^{(s)}(g)T^{(s)}_aT^{(s)}_b
              +\omega_{abc}U^{(s)}(g)T^{(s)}_c\}\right]\nonumber\\
&=&\sum_s\Tr\left[c^{(s)}(\rho)U^{(s)}(g)A^{(s)}(\rho)\right],
\lab{LapOnM}
\end{eqnarray}
where
\begin{equation}
A^{(s)}(\rho)=g^{ab}(\rho)T^{(s)}_aT^{(s)}_b
=\frac{1}{a^2(\rho)}T^{(s)}_1T^{(s)}_1
+\frac{1}{b^2(\rho)}T^{(s)}_2T^{(s)}_2
+\frac{1}{c^2(\rho)}T^{(s)}_3T^{(s)}_3.
\end{equation}
In (\ref{LapOnM}), we have used
the property of the spin connection $g^{ab}\omega_{abc}=0$.

The wave equation on $\cal M$ for the scalar field $\phi$ is
\begin{displaymath}
\left(E^2+\frac{1}{\sqrt g}\frac{\partial}{\partial\rho}
\sqrt g\frac{\partial}{\partial\rho}+\Delta^{\cal M}\right)\phi(\rho,g)=0,
\end{displaymath}
where $E^2=(p^0)^2-(p^1)^2-\cdots-(p^6)^2$
is the square of the momentum along the $M_7$ direction.
Expanding the field $\phi(r,g)$ by (\ref{expand}),
we get the equation for the coefficient $c^{(s)}(\rho)$:
\begin{equation}
\sum_s\left[\frac{1}{abc}\frac{d}{d\rho}abc\frac{d}{d\rho}
+E^2+A^{(s)}(\rho)\right]_{mk}c^{(s)}_{kn}(\rho)=0.
\lab{waveeq}
\end{equation}

If we have $a^2(\rho)=b^2(\rho)$, the matrix $A^{(s)}_{mn}$ is diagonalized.
\begin{equation}
A^{(s)}_{mn}
=\left[\frac{1}{a^2}s(s+1)
       +\left(\frac{1}{c^2}-\frac{1}{a^2}\right)m^2\right]\delta_{m,n}.
\end{equation}
For non-relativistic approximation,
we rewrite the energy $E$ in (\ref{waveeq}) as
\begin{equation}
E=M_{\rm D0}+H,\quad
M_{\rm D0}=\frac{n}{c_0},\quad
c_0\equiv c(\rho\rightarrow\infty),
\end{equation}
where $M_{\rm D0}$ is the D-particle mass,
and neglect the $H^2$ term to get
\begin{equation}
\left[H
+\frac{1}{2M_{\rm D0}}\left\{\frac{1}{abc}\frac{d}{d\rho}abc\frac{d}{d\rho}
+\frac{1}{a^2}s(s+1)\right\}
-\frac{c_0}{2}\left(\frac{1}{c^2}-\frac{1}{c_0^2}-\frac{1}{a^2}\right)n
\right]c_{nn}^{(s)}(\rho)=0.
\lab{nonrela}
\end{equation}
The potential between D-particle and 6-plane can be read off
from eqs.(\ref{nonrela}), (\ref{asym}) and (\ref{NDis}):
\begin{equation}
V(\rho)=\frac{c_0}{2}\left(\frac{1}{c^2}
            -\frac{1}{c_0^2}-\frac{1}{a^2}\right)n
       =N_D\frac{Q_M}{4\rho}+{\cal O}(\rho^{-2}),
\end{equation}
where $Q_M$ is magnetic charge of the 6-plane.
\begin{equation}
{\cal N}_{\rm TN} : Q_M=1,\quad
\ol{\cal N}_{\rm AH} : Q_M=-2,\quad
{\cal N}_{\rm AH} : Q_M=-4.
\end{equation}

\section{D-particle creation on an orientifold 6-plane}
In this section, we shall consider the gravitational wave and
the D-particle propagation
on the O6 or the O6+D6 background in the M-theory framework.
These backgrounds correspond to $\AH$ and $\ol\AH$, respectively.
On these manifolds, as we mentioned in the last section,
the $U(1)_z$ symmetry is broken.
This implies that the charge with respect to this symmetry
($=$D-particle number)
is not conserved near the 6-plane.
The purpose of this section is to study
this D-particle number changing process.
Unfortunately, we cannot solve the wave equation (\ref{waveeq}) exactly.
Therefore, we calculate the amplitude by means of numerical methods.

The orientifold flip $X^{7,8,9,11}\rightarrow-X^{7,8,9,11}$
changes the sign of the D-particle charge
and this fact prevents us from defining it globally.
Even if we consider the D-particle wave packet,
it convert into anti-D-particle when it goes around the 6-plane.
This D-particle number changing process is not
the one we mentioned above.
In the process we consider here,
not only the sign of the D-particle number
but also its absolute value changes.

In solving eq.(\ref{waveeq}),
we should take account of the projection
due to the discrete group $\Gamma\subset SU(2)_R$.
Namely, only $\Gamma$-invariant modes should be kept.
In the O6+D6 case, the group $\Gamma=Z_4$ is generated by $\exp(\pi T_x)$.
Because $\exp(\pi T_x)$ flips the sign of $m$,
the following condition should be satisfied.
\begin{equation}
c_{-n}(\rho)=c_n(\rho)\lab{flip}.
\end{equation}
(We omit the indices $s$ and $m$ of $c^{(s)}_{n,m}$.)
This implies that a state with a definite orbital angular momentum
is a superposition of a D-particle state and an anti-D-particle state.
In the O6 case, the group $\Gamma=D$ is generated
by two elements $\exp(\pi T_z)$ and $\exp(\pi T_x)$.
The matrix element of $\exp(\pi T_z)$ is
\begin{equation}
[\exp(\pi T_z)]_{nm}=(-1)^m\delta_{m,n}.
\end{equation}
Therefore, in addition to the condition (\ref{flip}),
we should impose the requirement that
$m$ is an even integer and $s$ is an integer.

Without calculation,
we can give some properties of the amplitude
$M^{O6/O6+D6}_{s,N_f\leftarrow N_i}$,
where $s$ is the orbital angular momentum
and $N_i$ and $N_f$ is the initial and the final D-particle number.
\begin{itemize}
\item
Eq.(\ref{waveeq}) does not contain Planck length $l_p$,
and the dimensional parameters contained in the amplitude are
$R_{11}$ (compactification radius) and $E$ (energy).
Therefore, the amplitude depends only on one parameter $R_{11}E$.
\item
In the high energy limit ($R_{11}E\gg1$),
the wave propagation is described by
geodesic orbits of a classical particle.
Therefore, the amplitude is independent of the wavelength
and it approaches to a constant.
\item
Because the wave equation (\ref{waveeq}) for O6 and that for O6+D6
are identical, we have
\begin{equation}
M^{O6+D6}_{s,2N_f\leftarrow2N_i}\left(\frac{R_{11}E}{2}\right)
=M^{O6}_{s,N_f\leftarrow N_i}(R_{11}E),
\lab{MandM}
\end{equation}
where eqs.(\ref{R11is}) and (\ref{NDis}) have been used.
\item
Because the non-zero elements of the matrix $A^{(s)}(\rho)$
in eq.(\ref{waveeq})
are only $A_{n,n}$ and $A_{n,n\pm2}$,
the change of $n$ is restricted to an even integer.
Using (\ref{NDis}), we get a selection rule for the D-particle number.
\begin{equation}
\Delta N_D=\frac{2}{p}\Delta n\in\frac{4}{p}\Z.\lab{DND}
\end{equation}
Namely, $\Delta N_D\in\Z$ for O6 and $\Delta N_D\in2\Z$ for O6+D6.
This is consistent with the relation (\ref{MandM}).
\end{itemize}

Hereafter, we focus on the case of $s=2$, which is the simplest case where
the D-particle number changing process occurs.
In this case, $c_n(\rho)$ has only two independent components
$c_0(\rho)$ and $c_{-2}(\rho)=c_2(\rho)$.
The metric of Atiyah-Hitchin manifold is
\begin{equation}
ds^2=(abc)^2d\eta^2+a^2(\rho)(\epsilon^1)^2+b^2(\rho)(\epsilon^2)^2+c^2(\rho)(\epsilon^3)^2,
\lab{AHmetric}
\end{equation}
where the functions $a$, $b$, and $c$ are given
in \cite{AH} in the parametric form
with parameter $0\leq\theta<\pi/2$:
\begin{equation}
ca=-2kk'^2K(k)\frac{dK(k)}{dk},\quad
bc=ca-2(k'K(k))^2,\quad
ab=ca+2(kK(k))^2,\quad
\eta=-\frac{K(k')}{\pi K(k)},
\end{equation}
where $k=\sin\theta$, $k'=\cos\theta$
and $K(k)$ is the complete elliptic integral.
The relation between the radial parameter $\eta$
and geodesic distance $\rho$
follows from (\ref{AHmetric}) as $d\rho=|abc|d\eta$.

\begin{figure}[hhh]
\epsfysize=5cm
\begin{center}
\leavevmode
\put(-10,57){$\frac{\pi}{\sqrt2}$}
\put(200,10){$\rho$}
\put(100,105){$|b(\rho)|$}
\put(100,75){$|a(\rho)|$}
\put(100,55){$|c(\rho)|$}
\epsfbox[0 360 300 560]{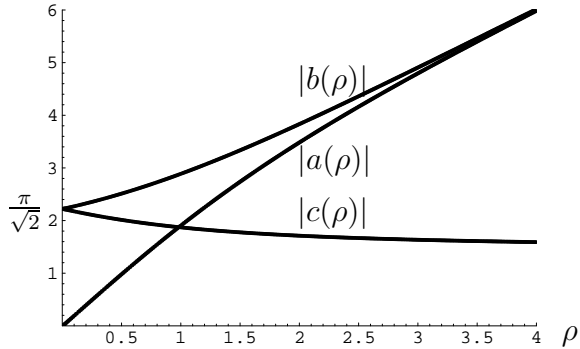}
\caption{Functions $a(\rho)$, $b(\rho)$ and $c(\rho)$
for ${\cal N}_{\rm AH}$.}
\lab{ahabc}
\end{center}
\end{figure}

Near the BOLT ($\rho=0$),
$\ol{\cal N}_{\rm AH}$(${\cal N}_{\rm AH}$)
is a plane bundle over $S^2$($RP^2$).
If we parameterize $\cal M$($=SU(2)$) by
\begin{equation}
g=e^{\phi T_x}e^{\theta T_z}e^{\psi T_x},\quad
0\leq\theta\leq\pi,\quad
0\leq\phi<2\pi,\quad
0\leq\psi<4\pi,
\end{equation}
then $(\theta,\phi)$ parameterizes
the base manifold $S^2$($RP^2$), and $(\rho,\psi)$
gives a polar coordinate on the plane fiber.
Since $b^2(\rho)\sim c^2(\rho)$ near the BOLT,
$U(1)_x$ is almost an isometry of ${\cal N}_{\rm AH}$
and it rotates the $\rho$-$\psi$ plane.
For the $s=2$ modes,
the eigenvalue of $T_x^2$ is $0$ (mode A) or $4$ (mode B).
The $\psi$-dependence of these modes are $\phi\sim{\rm const}$
and $\phi\propto\sin(2\psi+{\rm const})$, respectively.
To avoid singularity at $\rho=0$,
the $\rho$-dependence of these modes should be
$\phi\sim{\rm const}$ and $\phi\propto\rho$. 
(The Cartesian coordinate on the $\rho$-$\psi$ plane is given by
$(\rho\cos2\psi,\rho\sin2\psi)$ and the wave function of the mode B
is a linear function of this coordinate.)
Therefore if we adopt the generators
\newcommand{\sq}{\mbox{\tiny$\sqrt{\frac{3}{2}}$}}
\begin{equation}
T_x^{(2)}=\left(\begin{array}{ccccc}
0 & 1 \\
1 & 0 & \el{\sq} \\
& \el{\sq} & 0 & \el{\sq} \\
&& \el{\sq} & 0 & 1 \\
&&& 1 & 0
\end{array}\right),\quad
T_y^{(2)}=\left(\begin{array}{ccccc}
0 & -i \\
i & 0 & \el{-i\!\sq} \\
& \el{i\!\sq} & 0 & \el{-i\!\sq} \\
&& \el{i\!\sq} & 0 & -i \\
&&& i & 0
\end{array}\right),\quad
T_z^{(2)}=\left(\begin{array}{ccccc}
\el2 \\
& \el1 \\
&& \el0 \\
&&& \el{-1} \\
&&&& \el{-2}
\end{array}\right),
\end{equation}
the boundary conditions for these two modes at $\rho=0$ are
\begin{eqnarray}
\mbox{mode A} &:& c^2_0(\rho)\sim 1,\quad
                       c^2_{\pm2}(\rho)\sim-\sqrt{3/2},
\lab{Initial1}\\
\mbox{mode B} &:& c^2_0(\rho)\sim\sqrt{6}\rho,\quad
                       c^2_{\pm2}(\rho)\sim \rho.
\lab{Initial2}
\end{eqnarray}
Starting with the initial conditions (\ref{Initial1}) and (\ref{Initial2}),
we can calculate the wave functions
in the asymptotic region $\rho\rightarrow\infty$
numerically.
From its behavior,
we obtain the transition amplitude
between incoming gravitational wave and outgoing D-particle.
The result is shown in fig.\ref{result}.
\begin{figure}[hhh]
\epsfysize=7cm
\begin{center}
\leavevmode
\put(200,160){$|M^{O6}_{2,1\leftarrow0}|^2$}
\put(200, 70){$|M^{O6}_{2,0\leftarrow0}|^2$}
\put(320, 10){$R_{11}E$}
\put(45,205){threshold}
\epsfbox{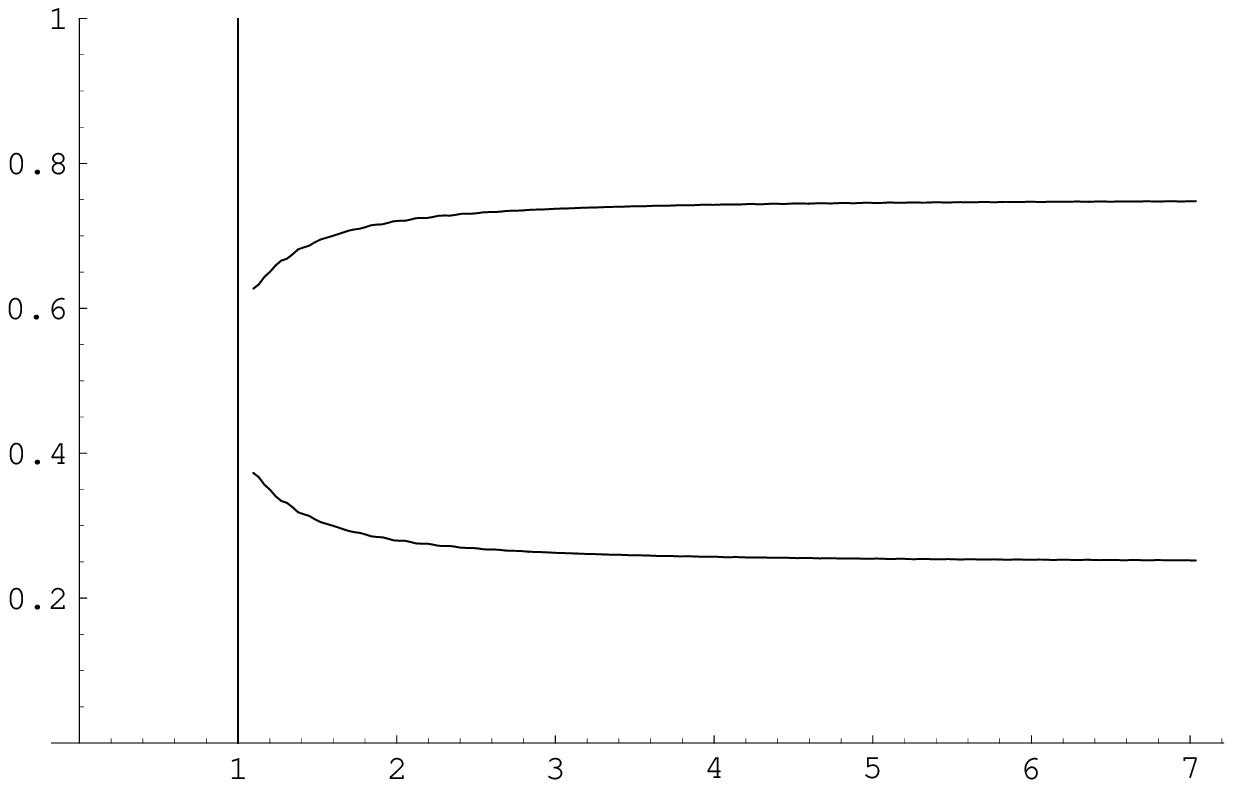}
\caption{Numerical results for the $s=2$ process.
D-particle creation amplitude $|M^{O6}_{2,1\leftarrow0}|^2$ vanishes
below the threshold ($R_{11}E<1$),
 and it approaches a constant ($\sim75\%$)
 in the classical region ($R_{11}E\gg1$).
This behavior is consistent with the expected properties.}
\lab{result}
\end{center}
\end{figure}

\vspace{4ex}
\noindent{\bf Acknowledgment}

I would like to thank H.\ Hata and T.\ Kugo
for valuable discussions and careful reading of the manuscript.

\end{document}